\documentstyle[12pt]{article}
\begin{document}
\baselineskip=0.81cm

\begin{titlepage}
\begin{center}
{\Large \bf  Analytic Solution of Non-resonant Multiphoton\\Jaynes-Cummings
Model with Dissipation\\
at Finite Temperature}
\end{center}

\begin{center}
{\bf    Le-Man Kuang}\\
{\small {\it   Theoretical Physics Dvision, Nankai Institute of
Mathematics,}}\\
{\small {\it   Tianjin 300071,People's Republic of China and}}\\
{\small {\it   Department of Physics and Institute of Physics, Hunan Normal
University,}}\\
{\small {\it   Hunan 410006, People's Republic of China}}\\
{\bf    Xin Chen}\\
{\small {\it   Institute of Acoustics and Department of Electrical Science and
Engineering,}}\\
{\small {\it   Nanjing University, Nanjing 210093, People's  Republic of
China}}\\
{\bf    Guang-Hong Chen and Mo-Lin Ge} \\
{\small {\it   Theoretical Physics Dvision, Nankai Institute of
Mathematics,}}\\
{\small {\it   Tianjin 300071,People's Republic of China }}\\
\end{center}

\begin{abstract}
\baselineskip=0.90cm
We derive a new master equation with and without Markovian approximation for
the reduced density operator of a quantum system. We study a multiphoton
Jaynes-Cummings
model (JCM)  with  the dissipation at a finite temperature by making use of
the new master equation under Markovian approximation. We present an exact
solution of a non-resonant  multiphoton JCM with the dissipstion by the use of
the superoperator technique, and obtain analytic expressions for the atomic
inversion and the intensity of the cavity field in the JCM.
It is shown that
the new master equation under the Markovian approximation can better describe
the suppression of the revivals of the inversion and oscillatary behavoirs of
 the photon-number distribution in the JCM. When the damping
vanishes, the usual expressions for the inversion and the intensity are
recovered.

\noindent PACS number(s): 42.50.Md, 42.50.Dv, 03.65.Bz
\end{abstract}
\hspace{0.5cm}
\newpage

\section{Introduction}
It is well known that the Jaynes-Cummings model (JCM) [1] is a model of
fundamental
theoretical importance, as the simplest nontrival model of two coupled
dissimilar quantum systems (the atom and the field). Without dissipation, it
is exactly solvable under the rotating-wave approximation. The solution in the
presence of dissipation is not only of theoretical interest, but also important
from a practical point of view since the dissipation would be always present in
any
experimental realization of the model. It was found that the collapses and
revivals of the inversion  oscillations predicted by the JCM are in agreement
with the experiments done with Rydberg atoms in a microwave cavity [2-4]. In
these experiments, the damping of the cavity mode is not negligbly small. Thus
for a detailed comparison with experiments, the effects of cavity losses must
be taken into account.

In the past few years, a number of authors have treated the JCM with the
dissipation by the use of analytic approximations [5,6] and numerical
calculations [7-11]. With the dissipation one has to solve a master equation
for the reduced density operator of the system.
It is well known that this
master equation is more difficult to solve. To our knowledge, no analytic
solution for this model has been given for general initial conditions and
temperature of the environment. Agarwal and Puri [12] present an analytic
solution for the initial state of the light field being a vaccum state.
Daeubler
{\it et al.} [13] found an analytic expression for the atomic inversion and the
intensity
of the field in the JCM with the dissipation when the reservoir is at zero
temperature.

In the present paper we derive a new master equation of the reduced density
operator for the system with and without Markovian approximation. This master
equation  differs from the previous one [5,6,14,15] due to different coupling
between the system and the reservoir and the presence of the renormalization
term in the total Hamiltonian. It enable us to discuss the influence
of the reservoir temperature on nonclassical effects in the JCM. We show that
the JCM is exactly sovable for the new master equation under the Markovian
approximation, and  analytic expressions of the atomic inversion and the
intensity
of the field  can be obtained for an arbitrary value of the damping within
the approximations used in deriving the master equation. When the damping
vanishes,
the usual results are recovered.

This paper is organized as follows: In Sec. II we derive the new master
equation
for the reduced density operator of the system with and without Markovian
approximation. In Sec.III we present an exact solution of the new master
equatiuon under the Markovian approximation for the non-resonant multiphoton
JCM where the field is initially in a coherent state and the atom in the
excited
state by the use of the superoperator technique. In Sec. IV we calculate the
atomic inversion and the intensity  of the field and discuss the limit of a
vanishing damping. We close with a summary.
\section{Master Equation}
We consider a system described a Hamiltonian $H$ and interacting with a
heat-bath environment (a reservoir) which consists of an infinite set of
harmonic oscillators. We assume that the system interacting with the
environmemt
can be described by the total Hamiltonian
\begin{equation}
H_T=H + \sum_{i}(\frac{p_i^2}{2m_i}+ \frac{1}{2}m_i\omega_i^2x_i^2)
+ \hbar H\sum_{i} C_ix_i +\hbar^2 H^2\sum_{i}\frac{|C_i|^2}{2m_i\omega_i^2}
\end{equation}
where
the second term  is the Hamiltonian of the reservoir,  the third one
 represents  the interaction  between the system and the reservoir
 with the coupling constant $\hbar C_i$, and the last one is the
renormalization term [16].
 Here we have adopted a simple coupling between the system and the reservoir
such
that it satisfies the condition $[V, H]=0$ which is of importance in the
back-action
evading and quantum-nondemolition schemes [17] and in the study of decoherence
of quantum system [18-20].

 For later convenience, we re write the total Hamiltoian (1) in the form
\begin{equation}
H_T=H + H_R + V
\end{equation}
where
\begin{equation}
H_R=\sum_i \hbar \omega_i b^+_ib_i,\hspace{2.0cm}
V=H\sum_{i=1}^{3}F_i
\end{equation}
with
\begin{equation}
F_1=\hbar \sum_iC_i b_i,\hspace{1.5cm}
F_2=\hbar \sum_iC_i b^+_i,\hspace{1.5cm}
F_3=\hbar^2 H\sum_{i}\frac{|C_i|^2}{\omega_i}
\end{equation}
where $b_i$, $b^+_i$ are the boson annihilation and creation operators for the
reservoir.

The total density operator for the system and the reservoir satisfies the
Liouville-von Neumann
equation in the interaction picture [15,21]
\begin{equation}
\frac{d\chi}{dt}=-\frac{i}{\hbar}[V(t), \chi]
\end{equation}
It is easy to that the above equation has the formal solution
\begin{equation}
\chi (t)=\chi (0)+\frac{1}{i\hbar}\int_{t_0}^{t} [V(t'), \chi (0)]\,{\rm d}t'
+(\frac{1}{i\hbar})^2 \int_{t_0}^{t} {\rm d}t' \int_{t_0}^{t'} {\rm d}t''
[V(t'), [V(t''), \chi(t'')]]
\end{equation}
Tracing both sides of the above equation over the reservoir, we obtain the
change in the reduced density operator $s=Tr_R \chi$ for the system
\begin{eqnarray}
s(t)-s(t_0)&=&\frac{1}{i\hbar}\int_{t_0}^{t} {\rm d}t'Tr_R [H\sum_i F_i (t'),
s(t_0)f(t_0)] \nonumber  \\
& &+ (\frac{1}{i\hbar})^2 \int_{t_0}^{t}{\rm d}t' \int_{t_0}^{t'}{\rm d}t''
Tr_R [V(t'), [V(t''), \chi(t'')]]
\end{eqnarray}
where  we have
assumed that the system and the reservoir are uncoupled and $\chi(0)$ can be
facterized as
$\chi(0)=s(t_0)f(0)$ at the initial time.

 We assume that the reservoir is initially in thermal equilibrium at a
temperature $T$.
Then it is straightforward to see that
\begin{equation}
\langle F_1\rangle_R=\langle F_2\rangle_R=0, \hspace{2.0cm}
\langle F_3\rangle_R=\Delta \omega'H
\end{equation}
where
\begin{equation}
\Delta \omega'=i\hbar \int_{0}^{\infty}{\rm d}\omega
\frac{J(\omega)|C(\omega)|^2}{\omega}
\end{equation}
where $J(\omega)$ is the spectral density of the reservoir.

 The total density operator [22,23] can be expressed as
\begin{equation}
\chi(t)=s(t)f(t) + \chi_c(t)
\end{equation}
where $\chi_c(t)$ represents the correlation between the system and the
reservoir
described by the density operator $f(t)$ at time $t$. The reservoir assumption
allows us to take $f(t)=f(0)$. Then, Eq.(7) becomes
\begin{eqnarray}
s(t)-s(t_0)&&=-(t-t_0)\Delta \omega'(H[H, s(t_0)]+[H, s(t_0)]H)\nonumber\\
&& +(\frac{1}{i\hbar})^2 \int_{t_0}^{t}{\rm d}t' \int_{t_0}^{t'}{\rm d}t''
\{ Tr_R [V(t'), [V(t''), s(t'')f(0)]]
+ Tr_R [V(t'), [V(t''), \chi_c(t'')]] \}
\end{eqnarray}
where the second term is a higher-order term than the first one and can be
neglected for the weak damping case [21]. After dropping the second term in
Eq.(6), differentiating
both sides of Eq.(11) with respect to time, we have the equation
\begin{eqnarray}\hspace{-0.5cm}
\partial s(t)\partial t=-\Delta \omega'(H[H, s(t_0)]+[H, s(t_0)]H) \nonumber \\
+(\frac{1}{i\hbar})^2 \int_{t_0}^{t}{\rm d}t'
 Tr_R [V(t), [V(t'), s(t')f(0)]]
\end{eqnarray}
where the $t'$ integration is over the correlation functions of the reservoir,
which are charecterized by a time which is short but finite.

Substituting the interaction term $V=\sum_i HF_i$ with a change of variable
$t'=t-\tau$ into Eq.(12), under Markovian approximation $s(t')=s(t)$, we
can rewrite Eq.(12) as
\begin{eqnarray} \hspace{-0.8cm}
{s}(t)&=&-\Delta \omega'(H[H, s(t_0)]+[H, s(t_0)]H) \nonumber \\
& &-[H, [H, s(t_0)]] \sum_{ij}\int_{0}^{\infty}{\rm d}\tau W_{ij}(\tau)
- [H,s(t_0)]H\sum_{ij}\int_{0}^{\infty}{\rm d}\tau [W_{ij}(\tau)-
W_{ji}'(\tau)]
\end{eqnarray}
where the correlation functions of the reservoir are defined by
\begin{eqnarray}
W_{ij}(\tau)&=&Tr_R F_i(t)F_j(t-\tau)f(0)=\langle F_i(t)F_j(t-\tau)\rangle _R
\\
W_{ij}'(\tau)&=&Tr_R F_i(t-\tau)F_j(t)f(0)=\langle F_i(t-\tau)F_j(t)\rangle _R
\end{eqnarray}
where all operators are in the interaction picture. These correlation functions
have been calculated in ref.[15].

It is not difficult to see that under the higher temperature approximation in
which the temperature is assumed high enough to so that the Markovian
approximation
is valid,
 Eq.(13) reduces to
\begin{equation}
{s}(t)=-\gamma [H, [H, s(t_0)]] - \Delta \omega [H, s(t_0)]H
\end{equation}
where $\gamma$ and $\Delta \omega $ are constants  which depends on the
temperature
and the spectral density of the reservoir, and given by
\begin{equation}
\gamma =\Delta \omega' +  \frac{kT}{\hbar}\lim_{\omega \rightarrow 0}
\frac{J(\omega)|C(\omega)|^2}
{\omega},
\end{equation}
\begin{equation}
 \Delta \omega= \Delta \omega' + 2i {\cal P} \int_{0}^{\infty}{\rm d} \omega
\frac{J(\omega)|C(\omega)|^2}{\omega}
\end{equation}
where ${\cal P}$ is the Cauchy principal part of the integration [15].


Eq.(16) is the master equation under the Markovian approximation in the
interaction picture for the system. It is easy to convert it to that in the
Schrodinger picture with this form:
\begin{equation}
\frac{d\rho _s(t)}{dt}=\frac{1}{i\hbar}[H, \rho(t)] - \gamma [H, [H, \rho(t)]]
- \Delta \omega [H, \rho(t)]
\end{equation}
which is the desired master equation for thr reduced density operator of
the system in the Schrodinger picture under the Markovian approximation.

Notice that in the derivation of the master equation (14) we have used the
 Markovian approximation $s(t')=s(t)$. Without use of the Markovian
approximation,
through the transformation of variable $t'=t-\tau$ in Eq.(12), then taking
the approximation $s(t-\tau)\doteq s(t)-\tau \partial s(t)\partial t$ where
$\tau$ is short
but finite.
To take the result under the Markovian approximation (16)
as $\partial s(t)\partial t$ here,
we then obtain the non-Markovian master equation for the reduced density
operator
of the system in the Schrodinger picture with the form
\begin{eqnarray}
\frac{d \rho _s}{dt}&=&\frac{1}{i\hbar}[H, \rho] - \gamma [H, [H, \rho ]] -
\Delta \omega [H, \rho ]H \nonumber \\
& &- \gamma \alpha [H, [H, [H, [H, \rho ]]]] - \Delta \omega \alpha [H, [H, [H,
\rho ]H]] \nonumber \\
& &- \gamma \beta [H, [H, [H, \rho]]]H - \Delta \omega \beta [H, [H, \rho ]]H -
\Delta \omega \beta [H, [H, \rho ]H]H
\end{eqnarray}
where $\gamma $ and $\Delta \omega $ are given by Eq.(17) and $\alpha$ and
$\beta $ are defined by
\begin{equation}
\alpha =\frac{2kT}{\hbar} {\cal P} \int_{0}^{\infty} {\rm d} \omega
\frac{J(\omega)|C(\omega)|^2}
{\omega^3}, \nonumber \hspace{1.0cm}
\beta =2\pi i \frac{d}{d\omega} [J(\omega)|C(\omega)|^2 ]\mid_{\omega =0}
\end{equation}

Comparison Eq.(20)  with Eq.(19) it is easy to see that the second and third
terms on the rhs of Eq.(20)
are the Markovian part, and the last four terms are the non-Markovian
contribution.
\section{Exact Solution for the JCM with Dissipation}
In this section we derive   an exact solution of the master equation under
the Markovian approximatiom for the non-resonant multiphoton JCM. If we
neglect the Lamb shift term in Eq.(14), the master equation under the
approximation becomes
\begin{equation}
\frac{d\rho (t)}{dt}=\frac{1}{i\hbar}[H, \rho (t)] - \gamma [H, [H, \rho(t)]]
\end{equation}
It is noticed that Eq.(22) has the same form as the Milburn's equation [24]
under diffusion
approximatiom, but they come from completely different physical mechanism. The
former originates from the dissipation while the latter is from the
uncontinuous
unitary evolution.

The non-resonant multiphoton Jaynes-Cummings Hammiltonian [25] describing an
interaction
of a two-level atom with a single-mode cavity field via an $m$-photon
process in the rotating-wave approximation is given by
\begin{equation}
\hat{H} = \hbar\omega (\hat{a}^+ \hat{a} + \frac{m}{2} \hat{\sigma}_3)
+\frac{\hbar }{2} (\omega_o -m\omega) \hat{\sigma}_3 +
\hbar \lambda (\hat{\sigma}_- \hat{a}^{+m} +\hat{\sigma}_+ \hat{a}^m),
\end{equation}
where $\omega$ is the frequency of the cavity field; $\omega_o$ is the atomic
transition frequency; $\lambda$ is the atom-field coupling constant;
$\hat{a}$ and $\hat{a}^+$ are the field annihilation and creation operators,
respectively; $\hat{\sigma}_3$ is the atomic-inversion operator and
$\hat{\sigma}_{\pm}$ are the atomic ``spin flip" operators which satisfy
the relations $[\hat{\sigma}_+ ,\hat{\sigma}_- ]=2\hat{\sigma}_3$ and
$[\hat{\sigma}_3 ,\hat{\sigma}_{\pm}]=\pm 2\hat{\sigma}_{\pm}$. For
simplicity, we take $\hbar =1$ throughout the paper.

We now start to find the exact solution for the density operator $\hat{\rho}
(t)$
of the master equation (22) applied to the Hamiltonian (23). Following the
approach in refs.[26,27]
 we introduce three auxiliary superoperators $\hat{R}$, $\hat{S}$
and $\hat{T}$ which are defined through their actions on the density operator,
respectively,
\begin{eqnarray}
\exp(\hat{R} \tau) \hat{\rho} (t)&=&\sum^{\infty}_{k=0} \frac{(2\gamma \tau)^k}
{k!} \hat{H}^k \hat{\rho} (t) \hat{H}^k    \\
\exp(\hat{S} \tau) \hat{\rho} (t)&=& \exp(-i\hat{H} \tau) \hat{\rho} (t)
\exp(i\hat{H} \tau)    \\
\exp(\hat{T} \tau) \hat{\rho} (t)&=&\exp(- \gamma \tau \hat{H}^2 )
\hat{\rho} (t) \exp(- \gamma \tau \hat{H}^2 )
\end{eqnarray}
where the Hamiltonia $\hat{H}$ is given by Eq.(23).

It is straightforward to obtain the formal solution of of the master equation
(22)
as follows:
\begin{equation}
\hat{\rho} (t)=\exp(\hat{R} t) \exp(\hat{S} t) \exp(\hat{T} t) \hat{\rho} (0)
\end{equation}
where $\hat{\rho} (0)$ is the density operator of the initial atom-field
system.

 We assume that initially the field is prepared in the coherent state
$\mid z\rangle $ defined by
\begin{equation}
\mid z\rangle = \sum^{\infty}_{n=0} \exp(-\frac{1}{2} \mid z \mid^2 )
\frac{z^n}{\sqrt{n!}} \mid n\rangle  \equiv \sum^{\infty}_{n=0} Q_n \mid
n\rangle
\end{equation}
where
\begin{equation}
Q_n = exp(-\frac{1}{2} |z|^2 ) \frac{z^n}{\sqrt n!}
\end{equation}
and the atom was prepared in its excited state $\mid e\rangle $, so that the
initial
density operator of the atom-field system takes this form:
\begin{equation}
\hat{\rho}(0)=  \pmatrix {
|z\rangle \langle z| & 0 \cr
0          & 0 \cr}
\end{equation}

We divide the Hamiltonian (23) into
a sum of two terms which commute with each other, i.e.,
\begin{equation}
\hat{H} =\hat{H}_o +\hat{H}_I,\hspace{0.3cm} [\hat{H}_o ,\hat{H}_I ]=0
\end{equation}
In the two-dimensional atomic basis $\hat{H}_0 $ and $\hat{H}_I$ take the
following form, respectively,
\begin{eqnarray}
&&\hat{H}_o = \omega \pmatrix{
\hat{n} +\frac{m}{2} & 0 \cr
0          & \hat{n}-\frac{m}{2} \cr}\nonumber\\
&&\hat{H}_I = \pmatrix{
\delta               &\lambda \hat{a}^m  \cr
\lambda \hat{a}^{+m}       & -\delta  \cr}
\end{eqnarray}
where $\hat{n}=\hat{a}^{\dagger}\hat{a}$ and  $\delta=\frac{1}{2}(\omega_0
-m \omega)$.

Similarly, the square of the Hamiltonian (23) can also be expressed as a sum
of two terms which commute with each other,
\begin{equation}
\hat{H}^2 =\hat{A} +\hat{B},\hspace{1.5cm} [\hat{A} ,\hat{B} ]=0
\end{equation}
where the representations of the operators $\hat{A}$ and $\hat{B}$ in the
two-dimensional atomic basis
take the following forms:
\begin{equation}
\hat{A} = \pmatrix{
\omega^2 (\hat{n} + \frac{m}{2} )^2  + \lambda^2 \hat{a}^m \hat{a}^{+m}
 + \delta^2    & 0 \cr
0                 & \omega^2 (\hat{n} - \frac{m}{2} )^2  +
 \lambda^2 \hat{a}^{+m} \hat{a}^m
 + \delta^2 \cr
}
\end{equation}
\begin{equation}
\hat{B} =  2 \omega
\pmatrix{
\delta (\hat{n} +\frac{m}{2})   &\lambda \hat{a}^m (\hat{n}- \frac{m}{2})\cr
\lambda (\hat{n} -\frac{m}{2}) \hat{a}^{+m}   & - \delta (\hat{n} -
\frac{m}{2}) \cr}
\end{equation}

It can be checked that operators $\hat{A}$ and $\hat{B}$ also commute with
$\hat{H}_o$ and $\hat{H}_I$ by the use of the following formulae:
\begin{equation}
\hat{a}^m \hat{a}^{+m} =\frac{(\hat{n} +m)!}{\hat{n}!}, \hspace{1.5cm}
\hat{a}^{+m} \hat{a}^{m} =\frac{\hat{n}!}{(\hat{n} -m)!},
\end{equation}
\begin{equation}
f(\hat{n}) \hat{a}^m =\hat{a}^m f(\hat{n} -m), \hspace{1cm}
\hat{a}^m f(\hat{n})=f(\hat{n} +m)\hat{a}^m
\end{equation}
\begin{equation}
f(\hat{n}) \hat{a}^{+m} =\hat{a}^{+m} f(\hat{n} +m), \hspace{1cm}
\hat{a}^{+m} f(\hat{n})=f(\hat{n} -m)\hat{a}^{+m}
\end{equation}
where $f(\hat{n})$ is an operator function on the number operator.

For convenience, we introduce the following auxiliary ``density"  operator:
\begin{equation}
\hat{\rho}_2 (t) =\exp(\hat{S} t)\exp(\hat{T} t) \hat{\rho} (0)
\end{equation}

{}From the definition of the superoperators and the  initial condition
(30), we find that
\begin{equation}
\hat{\rho}_2 (t) =\exp(-i \hat{H}_I t) \exp(-\gamma t \hat{B})
\hat{\rho}_1 (t) \exp(-\gamma t \hat{B})\exp(i\hat{H}_I t)
\end{equation}
where we have used
\begin{eqnarray}
\hspace{-0.6cm} \exp(-\gamma t \hat{A}) = \left( \begin{array}{cc}
e^{\{ -\gamma t [\omega^2 (\hat{n} +\frac{m}{2})^2 + \lambda^2
\hat{a}^m \hat{a}^{+m} + \delta^2] \}}     &  0   \\
0                          &  e^{\{ -\gamma t[\omega^2 (\hat{n} -\frac{m}{2})^2
+ \lambda^2
\hat{a}^{+m} \hat{a}^{m} + \delta^2] \}}
\end{array}  \right)
\end{eqnarray}

\begin{eqnarray}
\exp(-i \hat{H}_0 t) &=& \left( \begin{array}{cc}
exp[-i\omega t (\hat{n}+\frac{m}{2})]     &  0   \\
0                          &  exp[-i\omega t (\hat{n}-\frac{m}{2})]
\end{array}  \right)
\end{eqnarray}

The operator $\hat{\rho}_1 (t)$ in Eq.(40) is defined by
\begin{equation}
\hat{\rho}_1 (t)=\mid \Psi (t)\rangle \langle \Psi (t)\mid \otimes \mid
e\rangle \langle e\mid
\end{equation}
where
\begin{equation}
\mid \Psi (t)\rangle =\exp\{-\gamma t [\omega^2 (\hat{n} +\frac{m}{2})^2
+ \lambda^2 \hat{a}^m \hat{a}^{+m} + \delta^2] \} \mid ze^{-i\omega t}\rangle
\end{equation}
If we notice that the powers of the  operator $\hat{B}$
can be written as
\begin{eqnarray}
\hat{B}^{2k}&=& \left(  \begin{array}{cc}
A^k_+ (\hat{n})  & 0  \\
0 & A^k_- (\hat{n})
\end{array}  \right)
\end{eqnarray}
\begin{eqnarray}
\hat{B}^{2k+1} &=& \left(  \begin{array}{cc}
\delta (\hat{n}+\frac{m}{2})A^k_+ (\hat{n}) &
\lambda(\hat{n}+\frac{m}{2})A^k_+ (\hat{n}) \hat{a}^m \\
\lambda(\hat{n}-\frac{m}{2})A^k_- (\hat{n}) \hat{a}^{+m} &
- \delta (\hat{n}-\frac{m}{2})A^k_- (\hat{n})
\end{array}  \right)
\end{eqnarray}
where $k$ takes zero or an arbitrary positive integer and
\begin{equation}
A_+ (\hat{n})=(2\omega \delta)^2 (\hat{n} +\frac{m}{2})^2 + (2\lambda
\omega)^2 (\hat{n} +\frac{m}{2})^2 \hat{a}^m \hat{a}^{+m}
\end{equation}
\begin{equation}
A_- (\hat{n})=(2\omega \delta)^2 (\hat{n} -\frac{m}{2})^2 + (2\lambda
\omega)^2 (\hat{n} -\frac{m}{2})^2 \hat{a}^{+m} \hat{a}^{m}
\end{equation}
Then we can write the operator $\exp(-2\gamma t \hat{B})$ in the form
\begin{equation}
\exp(-2\gamma t \hat{B})= \left(  \begin{array}{cc}
E_+ (\hat{n},t)     &\hat{a}^m E_{3+}(\hat{n},t)\\
E_{3-}(\hat{n},t) \hat{a}^{+m}  & E_- (\hat{n},t)
\end{array}   \right)
\end{equation}
where
\begin{equation}
E_+ (\hat{n},t)=cos(\gamma t \sqrt{A_+ (\hat{n})}) -2\omega \delta
(\hat{n}+\frac{m}{2})
\frac{sinh(\gamma t \sqrt{A_+(\hat{n})})}{\sqrt{A_+ (\hat{n})}}
\end{equation}
\begin{equation}
E_- (\hat{n},t)=cos(\gamma t\sqrt{A_- (\hat{n})}) -2\omega \delta
(\hat{n}-\frac{m}{2})
\frac{sinh(\gamma t \sqrt{A_-(\hat{n})})}{\sqrt{A_- (\hat{n})}}
\end{equation}
\begin{equation}
E_{3+} (\hat{n},t)=-2\omega \lambda (\hat{n}-\frac{m}{2})
\frac{sinh(\gamma t \sqrt{A_+(\hat{n}-m)})}{\sqrt{A_+ (\hat{n}-m)}}
\end{equation}

\begin{equation}
E_{3-} (\hat{n},t)=-2\omega \lambda (\hat{n}-\frac{m}{2})
\frac{sinh(\gamma t \sqrt{A_-(\hat{n})})}{\sqrt{A_- (\hat{n})}}
\end{equation}
Similarly, we can write the operator $\exp(-i\hat{H}_I t)$ in the
two-dimensional atomic basis as
\begin{equation}
\exp(-i \hat{H}_I t)= \left(  \begin{array}{cc}
D_+ (\hat{n},t)        & D_{3+}(\hat{n},t) \hat{a}^m  \\
D_{3-}(\hat{n},t)\hat{a}^{+m}  & D_-(\hat{n},t)
\end{array}   \right)
\end{equation}
where
\begin{equation}
D_+ (\hat{n},t)=cos[t(\delta^2 +\lambda^2 \hat{a}^m \hat{a}^{+m})] -
\delta\frac{sin[t(\delta^2 +\lambda^2 \hat{a}^m \hat{a}^{+m})]}
{\sqrt{\delta^2 +\lambda^2 \hat{a}^m \hat{a}^{+m}}}
\end{equation}

\begin{equation}
D_- (\hat{n},t)=cos[t(\delta^2 +\lambda^2 \hat{a}^{+m} \hat{a}^{m})] -
\delta\frac{sin[t(\delta^2 +\lambda^2 \hat{a}^{+m} \hat{a}^{m})]}
{\sqrt{\delta^2 +\lambda^2 \hat{a}^{+m} \hat{a}^{m}}}
\end{equation}
and
\begin{equation}
D_{3+} (\hat{n},t)= -i \lambda
\frac{sin[t(\delta^2 +\lambda^2 \hat{a}^m \hat{a}^{+m})]}
{\sqrt{\delta^2 +\lambda^2 \hat{a}^m \hat{a}^{+m}}}
\end{equation}

\begin{equation}
D_{3-} (\hat{n},t)=-i \lambda
\frac{sin[t(\delta^2 +\lambda^2 \hat{a}^{+m} \hat{a}^{m})]}
{\sqrt{\delta^2 +\lambda^2 \hat{a}^{+m} \hat{a}^{m}}}
\end{equation}
For latter use, we list the following formulae
\begin{eqnarray}
\hspace{-0.8cm} \hat{a}^{+m} \left(  \begin{array}{c}
D_+ (\hat{n},t) \\ E_+ (\hat{n},t)
\end{array}  \right)
=\left( \begin{array}{c}
D_- (\hat{n},t) \\ E_- (\hat{n},t)
\end{array}    \right) \hat{a}^{+m},\hspace{0.5cm}
\hat{a}^{m} \left(  \begin{array}{c}
D_- (\hat{n},t) \\ E_- (\hat{n},t)
\end{array}  \right)
=\left( \begin{array}{c}
D_+ (\hat{n},t) \\ E_+ (\hat{n},t)
\end{array}    \right) \hat{a}^{m}
\end{eqnarray}
and
\begin{eqnarray}
\hspace{-0.8cm} \hat{a}^{m} \left(  \begin{array}{c}
D_{3-} (\hat{n},t) \\ E_{3-} (\hat{n},t)
\end{array}  \right)
=\left( \begin{array}{c}
D_{3+} (\hat{n},t) \\ E_{3+} (\hat{n},t)
\end{array}    \right) \hat{a}^{+m}, \hspace{0.2cm}
\hat{a}^{+m} \left(  \begin{array}{c}
D_{3+} (\hat{n},t) \\ E_{3+} (\hat{n},t)
\end{array}  \right)
=\left( \begin{array}{c}
D_{3-} (\hat{n},t) \\ E_{3-} (\hat{n},t)
\end{array}    \right) \hat{a}^{+m}
\end{eqnarray}

In the derivation of Eq.(54) we have used
\begin{eqnarray}
\hat{H_I}^{2k}=\left( \begin{array}{cc}
(\delta^2 +\lambda^2 \hat{a}^m \hat{a}^{+m})^k & 0 \\
0 & (\delta^2 +\lambda^2 \hat{a}^{+m} \hat{a}^{m})^k
\end{array} \right)
\end{eqnarray}
\begin{eqnarray}
\hat{H_I}^{2k+1}=\left( \begin{array}{cc}
\delta (\delta^2 +\lambda^2 \hat{a}^m \hat{a}^{+m})^k & \lambda (\delta^2
+\lambda^2 \hat{a}^m \hat{a}^{+m})^k \hat{a}^m \\
\lambda (\delta^2 +\lambda^2 \hat{a}^{+m} \hat{a}^{m})^k \hat{a}^{+m} & -\delta
(\delta^2 +\lambda^2 \hat{a}^{+m} \hat{a}^{m})^k
\end{array}        \right)
\end{eqnarray}
where $k$ takes zero or an arbitrary positive integer .

Then, from Eqs.(49) and (54) it follows that
\begin{eqnarray}
\exp(-i \hat{H}_I t)\exp(-\gamma t \hat{B} )= \left(  \begin{array}{cc}
F_+ (\hat{n},t)   &  F_{3+} (\hat{n},t)\hat{a}^m   \\
F_{3-} (\hat{n},t)\hat{a}^{+m}  & F_- (\hat{n},t)
\end{array}   \right)
\end{eqnarray}
where
\begin{equation}
F_+ (\hat{n}, t)=E_+ (\hat{n},t)D_+ (\hat{n},t) + E_{3+} (\hat{n},t)D_{3+}
(\hat{n},t) \hat{a}^m \hat{a}^{+m}
\end{equation}
\begin{equation}
F_- (\hat{n}, t)=E_- (\hat{n},t)D_- (\hat{n},t) + E_{3-} (\hat{n},t)D_{3-}
(\hat{n},t) \hat{a}^{+m} \hat{a}^{m}
\end{equation}
\begin{equation}
F_{3+} (\hat{n}, t)=E_{3+} (\hat{n},t)D_+ (\hat{n},t) + E_{+} (\hat{n},t)D_{3+}
(\hat{n},t)
\end{equation}
\begin{equation}
F_{3-} (\hat{n}, t)=E_{3-} (\hat{n},t)D_- (\hat{n},t) + E_{-} (\hat{n},t)D_{3-}
(\hat{n},t)
\end{equation}

Substituting Eqs.(43) and (63) into Eq.(40), we can obtain an
explicit expression  for the operator $\hat{\rho}_2 (t)$ as follows:
\begin{eqnarray}
\hat{\rho}_2 (t) =\left( \begin{array}{cc}
\hat{\Psi}_{11} (t)  &  \hat{\Psi}_{12} (t)  \\
\hat{\Psi}_{21} (t)  &  \hat{\Psi}_{22} (t)
\end{array} \right)
\end{eqnarray}
where we have used the  following symbol:
\begin{equation}
\hat{\Psi}_{ij} (t)= \mid \Psi_i (t)\rangle \langle  \Psi_j (t) \mid ,
\hspace{0.5cm} (i,j=1,2)
\end{equation}
with
\begin{equation}
\mid \Psi_1 (t) \rangle =F_+ (\hat{n},t) \mid \Psi (t)\rangle ,\hspace{1.0cm}
\mid \Psi_2 (t)=F_{3-} (\hat{n},t) \hat{a}^{+m} \mid \Psi (t)\rangle
\end{equation}
where $\mid \Psi (t)\rangle $ is given by equation (44).

Taking into account the definition of the superoperator $\hat{R}$,
we can obtain the action of the operator $\exp(\hat{R}t)$ on the ``density"
operator $\hat{\rho}_2 (t)$ as follows:
\begin{equation}
\hat{\rho} (t)=\sum^{\infty}_{k=0} (2\gamma t)^k \frac{1}{k!}
\hat{H}^k \hat{\rho}_2 (t) \hat{H}^k
\end{equation}
where the operator $\hat{H}$ and $\hat{\rho}_2 (t)$ are given by Eqs.(23)
and (68), respectively.

In fact,up to now we have  found the exact solution of the master equation (22)
for the
non-resonant multiphoton Jaynes-Cummings Hamiltonian (23) in the operator form
(71).
However,in most cases of practical interest, one needs to know the matrix
elements of the density operator $\hat{\rho} (t)$ in the two-dimensional atomic
basis to calculate expectation values of observables.
Although the form of the solution (71) is
pleasant, it is unconvenient in use.  Therefore in what follows we evaluate
matrix elements of the density operator in the two-dimensional atomic basis.

Since $\hat{H}_o$ commutes with $\hat{H}_I$, from equation (31) we obtain
\begin{eqnarray}
\hat{H}^k =\sum^{k}_{l=0} \left( \begin{array}{c} k\\l \end{array}
\right) \hat{H}^{k-l}_o \hat{H}^l_I
\end{eqnarray}

{}From Eq.(33) it follows that when $l$ takes an even number, we have
\begin{equation}
\hspace{-0.5cm} \hat{H}^{k-l}_{o} \hat{H}^l_I  =\frac{1}{2} \omega^{k-l}
\left( \begin{array}{cc}
(\hat{n}+\frac{m}{2})(\delta^2 +\lambda^2 \hat{a}^m \hat{a}^{+m})^{l/2}  & 0
\\
0  & (\hat{n}-\frac{m}{2})(\delta^2 +\lambda^2 \hat{a}^{+m} \hat{a}^{m})^{l/2}
\end{array}  \right)
\end{equation}
and when  $l$ is an odd number, we have
\begin{eqnarray}
\hspace{-1cm} \hat{H}^{k-l}_{o} \hat{H}^l_I =\frac{1}{2} \omega^{k-l}
\left( \begin{array}{cc}
\delta  (\hat{n}+\frac{m}{2})(\delta^2 +\lambda^2 \hat{a}^m
\hat{a}^{+m})^{(l-1)/2}
&\lambda (\hat{n}+\frac{m}{2})(\delta^2 +\lambda^2 \hat{a}^m
\hat{a}^{+m})^{(l-1)/2} \hat{a}^m \\
\lambda (\hat{n}-\frac{m}{2})(\delta^2 +\lambda^2 \hat{a}^{+m}
\hat{a}^{m})^{(l-1)/2} \hat{a}^{+m}
&-\delta  (\hat{n}-\frac{m}{2})(\delta^2 +\lambda^2 \hat{a}^{+m}
\hat{a}^{m})^{(l-1)/2}
\end{array}  \right)
\end{eqnarray}
Then,
\begin{eqnarray}
\sum_{l\in even} (\begin{array}{c} k\\l \end{array} )
\hat{H}^{k-l}_{o} \hat{H}^l_I  =\frac{1}{2}
\left( \begin{array}{cc}
\alpha^k_+ (\hat{n})+\alpha^k_- (\hat{n})  & 0    \\
0  & \beta^k_+ (\hat{n})+\beta^k_- (\hat{n})
\end{array}  \right)
\end{eqnarray}

\begin{eqnarray}
\sum_{l\in odd} (\begin{array}{c} k\\l \end{array} )
\hat{H}^{k-l}_{o} \hat{H}^l_I =\frac{1}{2}
\left( \begin{array}{cc}
\delta [\alpha^k_+ (\hat{n})-\alpha^k_- (\hat{n})]
& \lambda \frac{[\alpha^k_+ (\hat{n})-\alpha^k_- (\hat{n})]}{\sqrt{\delta^2
+\lambda^2 \hat{a}^m \hat{a}^{+m}}} \hat{a}^m \\
\lambda \frac{[\beta^k_+ (\hat{n})-\beta^k_- (\hat{n})]}{\sqrt{\delta^2
+\lambda^2 \hat{a}^{+m} \hat{a}^{m}}} \hat{a}^{+m}
&- \delta [\beta^k_+ (\hat{n})-\beta^k_- (\hat{n})]
\end{array}  \right)
\end{eqnarray}
where these operators $\alpha_{\pm}(\hat{n})$ and $\beta_{\pm}(\hat{n})$
are defined by, respectively,
\begin{equation} \hspace{-0.5cm}
\alpha_{\pm}(\hat{n})=\omega (\hat{n}+\frac{m}{2}) \pm \sqrt{\delta^2
+\lambda^2 \hat{a}^{m} \hat{a}^{+m}}, \\ \nonumber
\beta_{\pm}(\hat{n})=\omega (\hat{n}- \frac{m}{2}) \pm \sqrt{\delta^2
+\lambda^2 \hat{a}^{+m} \hat{a}^{m}}
\end{equation}
Substitution of Eqs.(75) and (77) into Eq.(72) yields that
\begin{eqnarray}
\hat{H}^k =\left( \begin{array}{cc}
\varphi^{(k)}_+ (\hat{n}) +\delta \varphi^{(k)}_- (\hat{n})
&\lambda\frac{\varphi^{(k)}_-(\hat{n})}{\sqrt{\delta^2 +\lambda^2 \hat{a}^{m}
\hat{a}^{+m}}} \hat{a}^m  \\
\lambda\frac{\phi^{(k)}_-(\hat{n})}{\sqrt{\delta^2 +\lambda^2 \hat{a}^{+m}
\hat{a}^{m}}} \hat{a}^{+m}
&\phi^{(k)}_+ (\hat{n}) -\delta \phi^{(k)}_- (\hat{n})
\end{array}  \right)
\end{eqnarray}
where
\begin{equation}
\varphi^{(k)}_{\pm} (\hat{n})=\frac{1}{2}
[\alpha^k_+ (\hat{n}) \pm \alpha^k_- (\hat{n})] , \hspace{1.0cm}
\phi^{(k)}_{\pm} (\hat{n})=\frac{1}{2}
[\beta^k_+ (\hat{n}) \pm \beta^k_- (\hat{n})]
\end{equation}
For convenience, we introduce
\begin{equation}
f^{(k)}_+ (\hat{n})=\varphi^{(k)}_+ (\hat{n}) +\delta \varphi^{(k)}_-
(\hat{n}),\nonumber
f^{(k)}_- (\hat{n})=\phi^{(k)}_+ (\hat{n}) -\delta \phi^{(k)}_- (\hat{n}
\end{equation}
Then, Eq.(78) becomes
\begin{eqnarray}
\hat{H}^k =\left( \begin{array}{cc}
f^{(k)}_+  (\hat{n})
&\lambda\frac{\varphi^{(k)}_-(\hat{n})}{\sqrt{\delta^2 +\lambda^2 \hat{a}^{m}
\hat{a}^{+m}}} \hat{a}^m  \\
\lambda\frac{\phi^{(k)}_-(\hat{n})}{\sqrt{\delta^2 +\lambda^2 \hat{a}^{+m}
\hat{a}^{m}}} \hat{a}^{+m}
&f^{(k)}_- (\hat{n})
\end{array}  \right)
\end{eqnarray}
Making use of equations (68) and (81), we can obtain the matrix
$\hat{\cal{M}}^{(k)} (t)\equiv \hat{H}^k \hat{\rho}_2 (t) \hat{H}^k$
where the matrix elements are explicitly given by
\begin{eqnarray} \hspace{-0.8cm}
\hat{\cal{M}}^{(k)}_{11} (t) &=& f^{(k)}_+ (\hat{n}) \hat{\Psi}_{11} (t)
f^{(k)}_+ (\hat{n})
 + \hat{a}^m \phi^{(k)'}_- (\hat{n})  \hat{\Psi}_{21} (t) f^{(k)}_+ (\hat{n})
 \nonumber  \\
& &+f^{(k)}_+ (\hat{n}) \hat{\Psi}_{12} (t) \phi^{(k)'}_- (\hat{n})
\hat{a}^{+m}
 + \hat{a}^m \hat{\phi}^{(k)'}_- (\hat{n}) \hat{\Psi}_{22} (t)\phi^{(k)'}_-
(\hat{n}) \hat{a}^{+m} \\
\hat{\cal{M}}^{(k)}_{22} (t) &=&\phi^{(k)'}_- (\hat{n})  \hat{a}^{+m}
\hat{\Psi}_{11} (t)
\hat{a}^m \phi^{(k)'}_- (\hat{n})
+ f^{(k)}_- (\hat{n}) \hat{\Psi}_{21} (t) \hat{a}^m \phi^{(k)'}_- (\hat{n})
\nonumber  \\
& &+ \phi^{(k)'}_- (\hat{n})  \hat{a}^{+m} \hat{\Psi}_{12} (m,t)f^{(k)}_-
(\hat{n})
+ f^{(k)}_- (\hat{n}) \hat{\Psi}_{22} (t) f^{(k)}_- (\hat{n} ) \\
\hat{\cal{M}}^{(k)}_{21} (t)&=& (\hat{\cal{M}}^{(k)}_{12} (t))^+   \nonumber
\\
&=& \phi^{(k)'}_- (\hat{n})  \hat{a}^{+m} \hat{\Psi}_{11} (t)
f^{(k)}_+ (\hat{n})
+  f^{(k)}_- (\hat{n}) \hat{\Psi}_{21} (m,t)  f^{(k)}_+ (\hat{n}) \nonumber  \\
& &+\phi^{(k)'}_- (\hat{n})  \hat{a}^{+m}  \hat{\Psi}_{12} (m,t) \phi^{(k)'}_-
(\hat{n}) \hat{a}^{+m}
+ f^{(k)}_- (\hat{n})\hat{\Psi}_{22} (m,t) \phi^{(k)'}_- (\hat{n})
\hat{a}^{+m}
\end{eqnarray}
with
\begin{equation}
\phi^{(k)'}_- (\hat{n})=\frac{\lambda}{\sqrt{\delta^2 + \lambda^2 \hat{a}^{+m}
\hat{a}^m}} \phi^{(k)}_- (\hat{n})
\end{equation}
 From Eqs.(71) and (82) to (85)  we finally can arrive at the explicit
expression
of the
 exact solution of the master  equation (22) for the non-resonant multiphoton
JCM  Hamiltonian (23).
With these results, one can further evaluate mean values of operators of
interest. In the
next section, we will use it to obtain analytic expressions of the atomic
inversion,
distribution of the photon number and the intensity of the field.

\section{Atomic Inversion and Intensity of the Field}
In this section, we derive analytic expressions of the atomic inversion and
the intensity of the field, and sdudy the influence of the dissipation
on nonclassical effects in the JCM.

The atomic inversion is defined as the probability of the atom being the
excited state minus the probability of being the ground state, that is
\begin{equation}
\langle \hat{\sigma}_3(t)\rangle =Tr[\hat{\rho}_A (t)\hat{\sigma}_3]
\end{equation}
where $\hat{\rho}_A(t)$ is the reduced desity operator of the atom, and can be
obtained
through tracing over the field part in (86).

Making use of the solution (86), it is easy to rewrite the inversion (87) as
\begin{equation}
\langle \hat{\sigma}_3(t)\rangle =\sum^{\infty}_{k,n=0} \frac{(2 \gamma
t)^k}{k!}
[\langle n|\hat{\cal M}^{(k)}_{11} |n \rangle - \langle n|\hat{\cal
M}^{(k)}_{22} |n \rangle ]
\end{equation}
It is not an easy task to calculate the two expectation values on the rhs of
Eq.(88).
{}From Eqs.(82) and (83) we get that
\begin{eqnarray} \hspace{-0.8cm}
\langle n\mid \hat{\cal{M}}^{(k)}_{11} (t)\mid n\rangle &=&(f^{(k)}_+ (n))^2
\mid\psi_1 (n,t)\mid^2  +(g^{(k)}(n+m))^2  \mid \psi_2 (n+m,t) \mid^2
\nonumber \\
& & +2Re \{ f^{(k)}_+ (n) g^{(k)}(n+m) \psi^*_1 (n,t) \psi_2 (n+m,t)\}  \\
\langle n\mid \hat{\cal{M}}^{(k)}_{22} (t) \mid n\rangle &=&(g^{(k)}(n))^2
\mid\psi_1 (n-m,t)\mid^2  +(f^{(k)}_- (n))^2 \mid \psi_2 (n,t) \mid^2
\nonumber  \\
& &+ 2Re \{f^{(k)}_ -(n) g^{(k)}(n) \psi_2 (n,t) \psi^*_1 (n-m,t)\}
\end{eqnarray}
where we have introduced the symbols:
\begin{equation}
g^{(k)}(n)=\sqrt{\frac{n!}{(n-m)!}} \phi^{(k)'}_- (n)
\end{equation}
\begin{equation}
\psi_i (n,t)= \langle n\mid \Psi_i (t)\rangle  \hspace{0.3cm} (i=1,2)
\end{equation}
Here $|\Psi_i (t)\rangle $ are given by Eq.(70). These functions
$f^{(k)}_{\pm}(n)$ and $\phi^{(k)'}_- (n)$
can be obtained through replacing the number operator $\hat{n}$ in the
operators $\hat{f}^{(k)}_{\pm}(n)$ and $\hat{\phi}^{(k)'}_- (n)$
by the number $n$.

Through a lengthy calculation Eqs.(89)and (90) can be written explicitly as
\begin{eqnarray} \hspace{-1.0cm}
\langle n|\hat{\cal M}^{(k)}_{11}(t)|n\rangle &=&
\frac{1}{4}\{ \alpha^{2k}_+(n) [(1+\delta )|F_+(n)|^2 +a_2(n)\frac{(n+m)!}{n!}
|F_{3-}(n+m)|^2 \nonumber \\
& &+2Re(F^*_+(n)F_{3-}(n+m))a(n)\sqrt{\frac{(n+m)!}{n!}}] \nonumber \\
& &+2\alpha^{k}_+(n) \alpha^k_-(n) [(1-\delta^2 )|F_+(n)|^2
-a_2(n)\frac{(n+m)!}{n!} |F_{3-}(n+m)|^2 \nonumber \\
& &+2Re(F^*_+(n)F_{3-}(n+m))\delta a(n)\sqrt{\frac{(n+m)!}{n!}}] \nonumber \\
& &+\alpha^{2k}_-(n) [(1-\delta^2 )|F_+(n)|^2 +a_2(n)\frac{(n+m)!}{n!}
|F_{3-}(n+m)|^2 \nonumber \\
& &-2Re(F^*_+(n-m)F_{3-}(n))(1+\delta )a(n)\sqrt{\frac{(n+m)!}{n!}}] \}|\Psi
(n,t)|^2
\end{eqnarray}

\begin{eqnarray} \hspace{-1.2cm}
\langle n|\hat{\cal M}^{(k)}_{22}(t)|n\rangle &=&
\frac{1}{4}\{ \beta^{2k}_+(n) [a_2(n-m)|F_+(n-m)|^2 +(1-\delta
)^2\frac{n!}{(n-m)!} |F_{3-}(n)|^2 \nonumber \\
& &+2Re(F^*_+(n-m)F_{3-}(n))a(n-m)(1-\delta )\sqrt{\frac{n!}{(n-m)!}}]
\nonumber \\
& &+2\beta^{k}_+(n) \beta^k_-(n) [a^2(n-m)|F_+(n-m)|^2 -(1-\delta^2)
|F_{3-}(n)|^2 \nonumber \\
& &+2Re(F^*_+(n-m)F_{3-}(n))\delta a(n)\sqrt{\frac{n!}{(n-m)!}}] \nonumber \\
& &+\beta^{2k}_-(n) [a^2(n)|F_+(n-m)|^2 +(1+\delta^2)|F_{3-}(n)|^2 \nonumber \\
& &-2Re(F^*_+(n-m)F_{3-}(n))(1+\delta )a(n)\sqrt{\frac{n!}{(n-m)!}}] \}|\Psi
(n-m,t)|^2
\end{eqnarray}
where the functions $\alpha_{\pm}$, $\beta_{\pm}$,$F(n)$ and $F_{3-}(n)$ are
given by replacing
the number operator $\hat{n}$ in the their corresponding operators by the
number $n$,
and we have introduced
\begin{eqnarray}
|\Psi (n,t)|^2&=&|\langle n|\Psi (t)\rangle |^2 \nonumber \\
& &=|Q_n|^2\exp\{-\gamma t[\omega^2(n+\frac{m}{2})^2 +\delta^2
+\lambda^2\frac{(n+m)!}{n!}]\}
\end{eqnarray}
\begin{equation}
a(n)=\sqrt{\frac{n! \lambda^2}{(n-m)! \delta^2 +n! \lambda^2 }}
\end{equation}

Substituting Eqs.(93) and (94) into Eq.(88) and taking into account the
relation
$\alpha^{k}_{\pm}(n)$=$\beta^{k}_{\pm}(n+m)$ we find that
\begin{eqnarray} \hspace{-1.2cm}
\langle \hat{\sigma}_3(t)\rangle &=&\frac{1}{4}\sum_{n=0}^{\infty}
|\Psi (n,t)|^2 \{|F_+(n)|^2 [(1+\delta^2 -a^2(n))\exp(2\gamma t\alpha^2_+(n))
\nonumber \\
& &+(2-2\delta^2 +a^2(n))\exp(2\gamma t\alpha_+(n)\alpha_-(n))
+((i-\delta)^2 -a^2(n))\exp(2\gamma t\alpha^2_-(n))] \nonumber \\
& &+|F_{3-}(n+m)|^2 [(a^2(n)-(1-\delta)^2) \frac{(n+m)!}{n!} \exp(2\gamma
t\alpha^2_+(n)) \nonumber \\
& &-2(1-\delta^2 +\frac{(n+m)!}{n!}a^2(n))\exp(2\gamma t\alpha_+(n)\alpha_-(n))
\nonumber \\
& & +a(n)(1+\delta+\sqrt{\frac{(n+m)!}{n!}}a^2(n))\exp(2\gamma t\alpha^2_-(n))]
\nonumber \\
& &+2Re(F^*_+(n)F_{3-}(n+m))a(n)\delta \sqrt{\frac{(n+m)!}{n!}} \nonumber \\
& &[\exp(2\gamma t\alpha^2_+(n)) +\exp(2\gamma t\alpha^2_-(n))] \}
\end{eqnarray}

As is well known, oscillations of the photon-number distribution in the JCM
is a kind of nonclassical effects of the cavity field. To see the influence
of the dissipation on this kind of nonclassical behavior, in what follows we
discuss photon statistics in the radiation field in the JCM.
   The reduced density operator of the cavity field can be obtained by taking
   the trace of the total density operator $\hat{\rho}(t)$ over the atomic
states,
that is, $\hat{\rho}_F=Tr_A\hat{\rho}(t)$.Then, the probability of finding $n$
photons in the radiation field is found to be
\begin{equation}
p(n,t)=\sum_{n=0}^{\infty} \frac{(2\gamma t)^k}{k!}
[\langle n|\hat{\cal M}^{(k)}_{11}(t)|n\rangle
+\langle n|\hat{\cal M}^{(k)}_{22}(t)|n\rangle]
\end{equation}
where the two expectation values on the rhs of Eq.(98) have been given
explicitly
in Eqs.(93) and (94).

Making use of Eq.(98), one can calculate easily the intensity of the cavity
field.
The result is
\begin{eqnarray} \hspace{-1.0cm}
\langle \hat{n} \rangle &=&\frac{1}{4}\sum_{n=0}^{\infty} n
|\Psi (n,t)|^2 \{|F_+(n)|^2 [(1+\delta^2 +a^2(n))\exp(2\gamma t\alpha^2_+(n))
\nonumber \\
& &+(2-2\delta^2 -a^2(n))\exp(2\gamma t\alpha_+(n)\alpha_-(n))
+((1-\delta)^2 +a^2(n))\exp(2\gamma t\alpha^2_-(n))] \nonumber \\
& &+|F_{3-}(n+m)|^2 [(a^2(n)+(1-\delta)^2) \frac{(n+m)!}{n!} \exp(2\gamma
t\alpha^2_+(n)) \nonumber \\
& &-2(1-\delta^2 +\frac{(n+m)!}{n!}a^2(n))\exp(2\gamma t\alpha_+(n)\alpha_-(n))
\nonumber \\
& &-a(n)(1+\delta-\sqrt{\frac{(n+m)!}{n!}}a^2(n))\exp(2\gamma t\alpha^2_-(n))]
\nonumber \\
& &+2Re(F^*_+(n)F_{3-}(n+m))a(n)\sqrt{\frac{(n+m)!}{n!}}
[(2-\delta)\exp(2\gamma t\alpha^2_+(n)) \nonumber \\
& &-4\delta\exp(2\gamma t\alpha_+(n) \alpha_-(n))
-2\exp(2\gamma t\alpha^2_-(n))] \}
\end{eqnarray}

In order to better understand the influence of the dissipation on nonclaasical
effects in the JCM, we consider the resonant case,i.e., $\delta =0$. In this
case,
the expressions of the inversion , photon-number distribution and the intensity
can simplified significantly.
The inversion reduces to
\begin{equation}
\langle\hat{\sigma}_3 (t)\rangle= \sum^{\infty}_{n=0} \mid Q_n \mid^2
\exp[-4\gamma \lambda^2 t
\frac{(n+m)!}{n!}] \cos[2\lambda t \sqrt{\frac{(n+m)!}{n!}}]
\end{equation}
The photon-number
distribution becomes
\begin{eqnarray}
p(n,t)&=& \frac{1}{2}|Q_n|^2 \left \{ 1+\exp[-4\gamma \lambda^2 t
\frac{(n+m)!}{n!}] \cos[2\lambda t
\sqrt{\frac{(n+m)!}{n!}}]  \right \}\nonumber \\
& &+\frac{1}{2}|Q_{n-m}|^2 \left \{ 1-\exp[-4\gamma \lambda^2t
\frac{n!}{(n-m)!}] \cos[2\lambda t
\sqrt{\frac{n!}{(n-m)!}}]  \right \}
\end{eqnarray}
And the intensity of the cavity field is given by
\begin{equation}
\langle \hat{n}(t)\rangle=\bar{n}
+\frac{m}{2}-\frac{m}{2}e^{-\bar{n}}\sum^{\infty}_{n=0}
\frac{\bar{n}^n}{n!} \exp[-4\gamma \lambda^2 t
\frac{(n+m)!}{n!}] \cos[2\lambda t \sqrt{\frac{(n+m)!}{n!}}]
\end{equation}
where $\bar{n}=|z|^2$ is the initial mean photon number in the
field.

 From the above expressions  we see that the damping term in Eq.(22)
leads to  the appearance of the decay factors
$\exp\{-4\gamma \lambda^2 t\frac{(n+m)!}{n!}\}$ and $\exp\{-4\gamma \lambda^2
t\frac{n!}{(n-m)!}\}$
in Eqs.(100), (101) and (102), which are responsible for the destruction of
revivals of the
atomic inversion  and the weakening of oscillatary behaviors of the
photon-number
distribution. In Fig.1 we plot the time evolution of the inversion. With
increasing
the damping $\gamma$, one can observe rapid deterioration of the revivals of
the
inversion.   It is worth mentioning that the inversion,
 oscillations of photon-number distribution and the intensity are weakened
with increasing temperature of the environment due to the relation (17).

\newpage

\vspace*{18cm}

\begin{center}
{\small FIG.1: The atomic inversion $\langle \hat{\sigma_3}\rangle $ as a
function of $t$ for $m=1, |z|^2=20$ and \\
(a) $\gamma =0.0001$, (b) $\gamma =0.0005$, (c) $\gamma=0.001$. Here we have
let
$\lambda =1$}
\end{center}

It is obvious that when the damping vanishes, Eqs.(100), (101) and (102) take
the form
 \begin{equation}
\langle\hat{\sigma}_3 (t)\rangle= \sum^{\infty}_{n=0} \mid Q_n \mid^2
 \cos[2\lambda t \sqrt{\frac{(n+m)!}{n!}}]
\end{equation}
 \begin{equation}
p(n,t)=|Q_n |^2 \cos^2[\lambda t\sqrt{\frac{(n+m)!}{n!}}]
+|Q_{n-m}|^2 \sin^2[\lambda t\sqrt{\frac{(n+m)!}{n!}}]
\end{equation}
and
\begin{equation}
\langle \hat{n}(t)\rangle=\bar{n}
+\frac{m}{2}-\frac{m}{2}e^{-\bar{n}}\sum^{\infty}_{n=0}
\frac{\bar{n}^n}{n!} \cos[2\lambda t \sqrt{\frac{n!}{(n-m)!}}]
\end{equation}
which are just the usual expressions without dissipation.

\section{Summary}
we have derived a new master equation with and without the Markovian
approximatiom
for the reduced density operator of a quantim system.
Making use of the new master
equation under the Markovian approximation, we we have studied the JCM with the
 dissipation at a finite temperature. We have  found an analytic solution of
the
 non-resonant multiphton JCM with the dissipation  and   obtained analytic
expressions
of the atomic inversion, the photon-number distribution and the intensity of
the
cavity field,which can reduce to usual expressions when the damping vanishes.
We have shown that the damping suppresses the revivals of the atomic inversion
and the oscillatary behaviors of the photon-number distribution. It must be
mentioned that  we do not discuss applications of the non-Markovian
equation in the present paper, but it may be useful  for the study of  the
non-Markovian relaxation processes in transient optical phenomena [28].

We would like to thank  our colleagues at Theoretical Physics Division , Nankai
Institute of Mathmatics for their  useful discussions
and comments. This research was supported by the National Natural Science
Foundation of China.

\stop
\end{document}